\def\spose#1{\hbox to 0pt{#1\hss}}

\def\multleft#1{\hbox to size{\vbox {\halign {\lft{##}\cr #1}}\hfill}\par}
\def\multright#1{\hbox to size{\vbox {\halign {\rt{##}\cr #1}}\hfill}\par}

\def\today{\ifcase\month\or January\or February\or March\or April\or May\or
      June\or July\or August\or September\or October\or November\or December\fi
      \space\number\day, \number\year}
\def\s{\hbox{\phantom{5}}}	

\def\km{{\rm\thinspace km}}

\def\Mpc{{\rm\thinspace Mpc}}

\def\s{{\rm\thinspace s}}

\def\kmps{\hbox{$\km\s^{-1}\,$}}

\def\kmpspMpc{\hbox{$\kmps\Mpc^{-1}$}}

\def\H2{\hbox{H$_{2}$}}

\newcommand{\gtsim}{\mbox{{\raisebox{-0.4ex}{$\stackrel{>}{{\scriptstyle\sim}}
$}}}}
\newcommand{\ltsim}{\mbox{{\raisebox{-0.4ex}{$\stackrel{<}{{\scriptstyle\sim}}
$}}}}
\documentclass[usegraphicx]{mn2e}

\usepackage{times}
\usepackage{amssymb}
\include{defn}

\begin{document}
\hsize=6truein

\title[The $M_{bh} - L_{\rm rad}$ relation for flat-spectrum radio-loud quasars]{On the black-hole mass -- radio
luminosity relation for flat-spectrum radio-loud quasars}
\author[M.J.~Jarvis  \& R.J.~McLure ]
{Matt J.~Jarvis$^{1}$\thanks{Email: jarvis@strw.leidenuniv.nl} \& Ross J.~McLure$^{2}$ \\
\footnotesize
$^{1}$Sterrewacht Leiden, Postbus 9513, 2300 RA Leiden, The Netherlands. \\
$^{2}$Institute for Astronomy, University of Edinburgh, Royal Observatory, Edinburgh EH9 3HJ \\}
\maketitle

\begin{abstract}
A new analysis of the connection between black-hole mass and radio
luminosity in radio-selected flat-spectrum quasars (FSQ) is
presented. In contrast to recent claims in the literature, we find no
evidence that the black-hole masses of radio-selected FSQ are
systematically lower that those of luminous optically-selected
radio-loud quasars. The black-hole masses of the FSQ are estimated via
the virial black-hole mass estimator which utilizes the line-width of
the H$\beta$ emission line as a tracer of the central gravitational
potential. By correcting for the inevitable effects of inclination,
incurred due to the FSQ being viewed close to the line of sight, we
find that the black-hole masses of the FSQ with intrinsically powerful radio jets are confined,
virtually exclusively, to $M_{bh} >10^{8}$~M$_{\odot}$. This is in good
agreement with previous studies of optically selected FSQ and
steep-spectrum radio-loud quasars. 

Finally, following the
application of a realistic Doppler boosting correction, we find that the FSQ occupy a wide range in intrinsic radio luminosity, and that many sources would be more accurately classified as radio-intermediate or radio-quiet quasars.
This range in radio luminosity suggests that the FSQ are fully consistent with an upper boundary on radio power of
the form $L_{\rm 5GHz} \propto M_{bh}^{2.5}$.
\end{abstract}
\begin{keywords}
galaxies:active - galaxies:nuclei - quasars:general - 
radio continuum:galaxies - quasars:emission lines 
\end{keywords}

\section{INTRODUCTION}

It is now widely believed that the energy emitted by active galactic
nuclei (AGN) is a consequence of mass accretion onto a central
supermassive black hole. It is possible that the mass of the central
black hole may be a crucial fundamental parameter in understanding the
physics of AGN.

In particular, one question which has recently received a great deal of
attention in the literature is whether or not the mass of an AGN's black
hole is strongly related to it's radio luminosity. This question is of
importance, because if it is established that radio-loud and
radio-quiet quasars have different black-hole mass distributions, it
may help explain why quasars of comparable optical luminosities can
differ in their radio luminosity by many orders of magnitude. On the
contrary, if radio-loud and radio-quiet quasars are found to have
essentially identical black-hole mass distributions, then the search
for the origin of radio loudness must move to some other physical
parameter such as black-hole spin.

Recent studies of the black-hole mass -- radio luminosity relation 
($M_{bh}-L_{\rm rad}$) relation have produced apparently contradictory
results. Franceschini,
Vercellone \& Fabian (1998) found that high frequency (5~GHz) radio
luminosity correlated strongly with the mass of the central black hole in a
sample of nearby inactive galaxies from the work of Kormendy \& Richstone
(1995). This surprisingly tight correlation had the form $L_{\rm 5GHz}
\propto M_{bh}^{2.5}$, which they proposed was indicative of advection
dominated accretion being the primary mechanism in controlling the radio
output of objects with a low accretion rate.

Using the spectral data of Boroson \& Green (1992), Laor (2000)
investigated the relation between black-hole mass and radio
luminosity in the Palomar-Green quasar sample using the virial
black-hole mass estimator. The results from this 
analysis pointed to an apparent bi-modality in black-hole mass, 
with virtually all of the radio-loud quasars containing 
black holes with masses $M_{bh} > 
10^{9}~$M$_{\odot}$, whereas the majority of quasars with black hole
masses $M_{bh} < 3 \times 10^{8}~$M$_{\odot}$ were radio quiet.

A similar result was arrived at by McLure \& Dunlop (2002) using a
sample of radio-loud and radio-quiet quasars matched in terms of both
redshift and optical luminosity. The results of this study indicated
that the median black-hole mass of the radio-loud quasars was a factor
of $\sim2$ larger than that of their radio-quiet counterparts. However, the
substantial over-lap between the black-hole mass distributions of the
two quasar samples indicated in addition that black-hole mass could
not be the sole parameter controlling radio power.

The FIRST Bright Quasar Survey (FBQS; Gregg et al. 1996; White et 
al. 2000) has also been used to probe the relationship between radio 
luminosity and black-hole mass. The radio-loud quasars in the FBQS 
fill the gap between radio-quiet and radio-loud quasars and are thus 
ideal probes of medium power radio sources. From their study of
the FBQS, Lacy et al. (2001) found a continuous variation of 
radio luminosity with black hole mass, 
in addition to evidence supporting the view that radio power also 
depends on the accretion rate relative to the Eddington limit.

However, two studies have recently questioned whether there is any
real connection between black-hole mass and radio power. 
Using a compilation of objects which ranged from nearby quiescent galaxies
to low-redshift quasars, Ho (2002) re-examined the relationship between
black-hole mass and radio luminosity. In contrast to the studies
outlined above, the results of this analysis suggested that there was
no clear relationship between radio power and black-hole mass, leading
the author to conclude that radio-loud AGN could be powered by black holes with a large
range of masses ($10^{6}\rightarrow
10^{9}{\rm M}_{\odot}$).

At least part of the disagreement about what black-hole mass is
required to produce a radio-loud AGN is due to different methods of
classifying what constitutes `radio-loud'. For example, the
classification adopted by Ho (2002) is the so-called radio-loudness
parameter ${\mathcal R}$, the ratio of radio-to-optical luminosity, under which
any object satisfying ${\mathcal R}>10$ is classified as radio-loud. Although
there can be little doubt that objects with black-hole masses of 
$10^{6}\rightarrow 10^{8}{\rm M}_{\odot}$ can be radio-loud under 
this classification, it is worth noting that only one object in the
sample of Ho (2002) with an estimated black-hole mass of 
$<10^{8}{\rm M}_{\odot}$, would also satisfy the alternative 
radio-loudness criterion of $L_{\rm 5GHz}>10^{24}$W~Hz$^{-1}$sr$^{-1}$ 
(Miller et al. 1990).

Following their study of the black-hole masses and host-galaxy
properties of low redshift radio-loud and radio-quiet quasars, Dunlop
et al. (2002) proposed an alternative view of the $M_{bh}-L_{\rm rad}$ plane. 
They argue that the location of both active and
non-active galaxies on the $M_{bh}-L_{\rm rad}$ plane appears to be 
consistent with the existence of an 
upper and lower envelope, both of the approximate form $L_{\rm 5GHz}
\propto M_{bh}^{2.5}$, but separated by some 5 orders of magnitude in
radio power. In this scheme the upper and lower envelopes delineate 
the maximum and minimum radio luminosity capable of being produced by 
a black hole of a given mass. Dunlop \& McLure (2002) demonstrate that
this scenario naturally describes the FBQS data of Lacy et
al. (2001), and we note here that it is also consistent with the
findings of Ho (2002).

However, the recent study by Oshlack, Webster \& Whiting 
(2002; hereafter OWW02) of the black-hole masses of a sample
of flat-spectrum radio-loud quasars from the Parkes Half-Jansky Flat
Spectrum sample of Drinkwater et al. (1997), casts doubt on the
existence of any upper threshold in the $M_{bh}-L_{\rm rad}$ plane. OWW02
found that their flat-spectrum quasars, which are
securely radio-loud with respect to either classification mentioned
above, harbour black-hole masses in the range $10^{6}\rightarrow
10^{9}M_{\odot}$, and therefore lie well above the upper $L_{\rm 5GHz}
\propto M_{bh}^{2.5}$ boundary proposed by Dunlop et al. (2002). The conclusion
reached by OWW02 following this result was that previous studies have
actively selected against including powerful radio sources with
relatively low black-hole masses, due to their concentration on luminous,
 optically selected radio-loud quasars.

In this paper we use the OWW02 sample to re-examine the position of
these flat-spectrum radio-loud objects on the $M_{bh}-L_{\rm rad}$ plane 
when both Doppler boosting effects and 
the likely geometry of the broad-line region are taken into
account. The motivation behind this re-analysis is to determine, once
the inevitable complications associated with flat-spectrum quasars
are considered, whether this sample of objects does genuinely violate the
apparent upper boundary to radio luminosity suggested by previous studies.

The paper is set out as follows, in Section~\ref{sec:virial} we briefly summarise the main aspects of the virial black-hole mass estimate. In Section~\ref{sec:sample} we provide a brief
description of the OWW02 sample. In Section~\ref{sec:doppler} we discuss
the amount of Doppler boosting one might expect in flat-spectrum radio
sources, and in Section~\ref{sec:reanalysis2} we consider what effect a BLR with a flattened disk-like geometry will have on the virial black-hole mass estimates. In Section~\ref{sec:reanalysis} we briefly discuss how the combination of these may affect the $L_{\rm rad} - M_{bh}$ relation for flat-spectrum quasars.The implications of this work are discussed in 
Section~\ref{sec:implications}.  All cosmological calculations 
presented in this paper assume $\Omega_{\rm M} = 0.3$,
$\Omega_{\Lambda} = 0.7$, $H_{\circ} = 70$~\kmpspMpc.

\begin{figure}
\includegraphics[width=0.48\textwidth,angle=0]{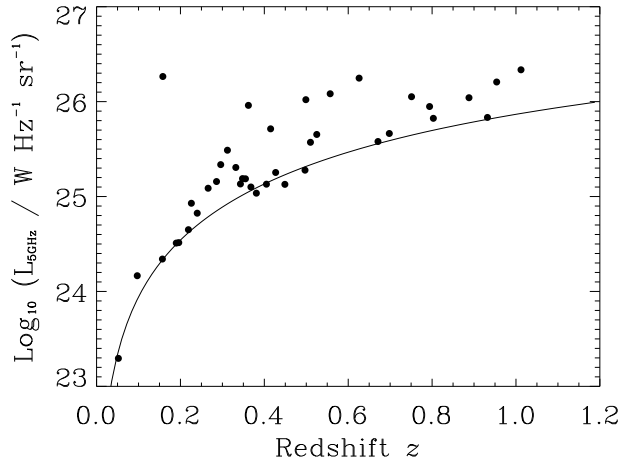}
\caption{Rest-frame 5~GHz radio luminosity versus redshift for the flat-spectrum sources analysed by OWW02. The solid line represents the flux-density limit of the Parkes half-Jy flat-spectrum sample assuming a spectral index of $\alpha = 0.5$. Some objects appear below this line because the remeasured spectral index using the whole radio spectrum is steeper than $\alpha = 0.5$  }
\label{fig:pzplane}
\end{figure}

\section{Measuring Black hole masses}\label{sec:virial}
In recent years a number of methods have been utilized to 
estimate the mass of the black holes in both active and 
non-active galaxies.  The two methods which have received most
attention are based on the correlations found between black-hole mass 
and both the luminosity of the host galaxy spheroid component 
(e.g. Kormendy \& Richstone 1995; Magorrian et al. 1998), and the 
stellar velocity dispersion (Ferrarese \& Merritt 2000, Gebhardt et al. 2000).

However, with reference to determining the masses of {\it active}
black-holes, the virial black-hole mass estimator 
has now become well established.  The virial black-hole mass 
estimate uses the width of the broad Hydrogen Balmer emission 
lines to estimate the broad line region (BLR) velocity dispersion. The 
black hole mass may then be calculated under the assumption that 
the velocity of the BLR clouds is Keplerian:
\begin{equation}
M_{bh} = R_{\rm BLR} V^{2} G^{-1},
\end{equation}
where $V$ is the velocity dispersion of the BLR clouds, usually 
estimated from the full-width half maximum (FWHM) of the H$\beta$ line, 
and $R_{\rm BLR}$ is the radius of the broad-line region.

The measurement of the radius of the BLR is ideally achieved by 
reverberation mapping, in which the continuum and line variations of a
number of sources are monitored over a number of years. The time lag
between the continuum variations and the variations of broad lines
will thus give the distance of the line emitting clouds from the
ionizing source (e.g. Wandel, Peterson \& Malkan 1999; Kaspi et al. 2000).

Unfortunately, reverberation mapping of quasars is extremely time
consuming and it remains unrealistic that the black-hole masses of a
large sample of quasars can be measured in this way. However, the
existing reverberation mapped data does provide an alternative.  The
radius of the BLR is found to be correlated with the monochromatic AGN
continuum luminosity at 5100\AA, $\lambda L_{5100}$ 
(e.g. Kaspi et al. 2000). Therefore, this correlation can be 
exploited to produce a virial black-hole mass estimate from a 
single spectrum covering the H$\beta$ emission line.
 
In this paper we adopt the calibration of the correlation between 
$R_{\rm BLR}$ and $\lambda L_{5100}$ from McLure \& Jarvis (2002), i.e.
\begin{equation}
R_{\rm BLR} = (25.4 \pm 4.4)[\lambda L_{5100} / 10^{37}{\rm W}]^{(0.61
\pm 0.10)},
\end{equation}
which, when combined with BLR velocity estimate from the $H\beta$
FWHM, leads to a black-hole mass estimate given by :
\begin{equation}
\frac{M_{bh}}{\rm M_{\odot}} = 4.74 \left( \frac{\lambda L_{5100}}{10^{37} {\rm W}} \right )^{0.61} \left( \frac{{\rm FWHM}(H\beta)}{\rm km~s^{-1}}\right)^{2},
\end{equation}
\noindent
which is adopted throughout this paper. We note that using the black-hole
mass estimator based on the $R_{\rm BLR}-\lambda L_{5100}$ relation of
Kaspi et al. (2000), as adopted by OWW02, results
in a typical difference in estimated black-hole mass of $\ltsim
0.1$~dex when converted to our chosen cosmology, 
and therefore has little bearing on the conclusions of this paper.
\begin{figure}
\includegraphics[width=0.48\textwidth,angle=0]{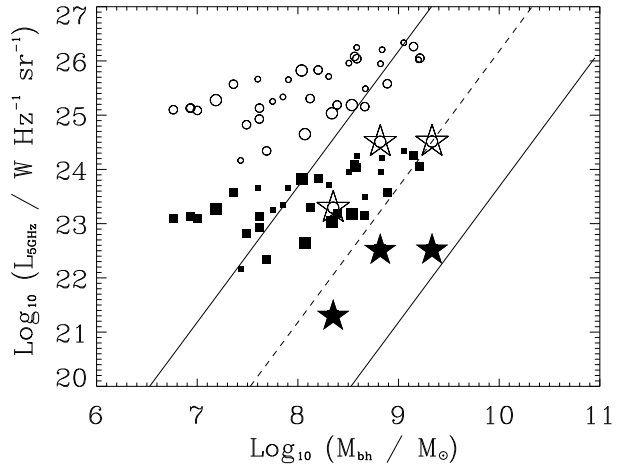}
\caption{Total radio luminosity at 5~GHz ($L_{\rm 5GHz}$) versus black-hole mass ($M_{bh}$) for the flat-spectrum quasars from OWW02. The open circles are the original points of OWW02 converted to our cosmology without a correction for Doppler boosting. The filled squares are the same sources but with the flux-density decreased by a factor of $\sim 100$, in accordance with what we expect the average Doppler boosting factor to be. The size of the symbols are scaled with radio spectral index, the smallest symbols represent $\alpha < 0.0$; medium sized symbols represent $0.0 < \alpha < 0.5$; large symbols represent $\alpha > 0.5$. The stars represent the three sources with $\alpha > 0.5$ discussed in section~\ref{sec:reanalysis}. 
 The lines are relations of the form $L_{\rm 5~GHz} \propto
 M_{bh}^{2.5}$ offset by 2.5 orders of magnitude from each other and
 represent the envelopes discussed in Dunlop et al. (2002), converted to
 our chosen cosmology. As can be seen, the de-boosting of the radio
 flux density in these sources is sufficient to push the majority of
 the sources within the region enclosed by the upper and lower lines.}
\label{fig:radpowvbhm1}
\end{figure}

\section{The flat-spectrum sample}\label{sec:sample}
The data used in this analysis are derived from the Parkes Half-Jansky
Flat Spectrum sample of Drinkwater et al. (1997), where full details
may be found. The crucial selection criteria for this work are a flux
limit at 2.7~GHz of $S_{2.7} > 0.5$~Jy and a spectral index between
2.7~GHz and 5~GHz $\alpha_{2.7}^{5.0} < 0.5$, where $S_{\nu} \propto
\nu^{-\alpha}$. The total sample comprises 323 radio sources. A
sub-sample of 39 of these sources was compiled by OWW02 for which it
was possible to obtain spectra covering the broad H$\beta$ 
emission line from various sources in the literature (Francis et
al. 2001; Wilkes et al. 1983). 

 We have remeasured all of the radio spectral indices for the 39
sources in this sub-sample using all of the publicly available radio
data. Consequently, our rest-frame 5~GHz radio luminosities differ slightly
to those of OWW02. It is worth noting at this point that some of the
`flat-spectrum' objects in this sample actually have spectral indices steeper
than the spectral index cut-off of $\alpha =0.5$. This may be due to a
number of uncertain effects, including variability of the source and any
spectral curvature. Fig.~\ref{fig:pzplane} shows the radio
luminosity redshift plane for the OWW02 sample after converting to our
chosen cosmology.

\begin{figure}
\includegraphics[width=0.48\textwidth,angle=0]{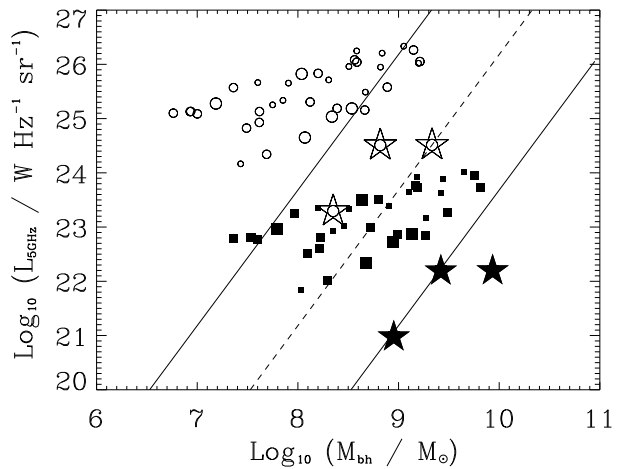}
\caption{As Fig.~\ref{fig:radpowvbhm1} but with the filled symbols representing sources now displaced to the right of the plot due to including a mean correction to the FWHM of the broad Balmer lines, in accordance with the disk-BLR model of McLure \& Dunlop (2002).
 As can be seen, all sources may now easily reside within the upper and lower boundaries represented  by the solid lines. The outliers to the right of the lower boundary are easily reconciled with the envelope as the correction for the disk-like geometry is probably an overestimate for these sources where the data suggests that they are not beamed directly along out line-of-sight. Thus, these sources would move up and to the left of the plot if this is indeed the case.}
\label{fig:radpowvbhm}
\end{figure}

\section{Doppler Boosting of the radio flux in flat-spectrum radio sources}\label{sec:doppler}

Flat-spectrum radio samples unavoidably contain a mix of radio source
populations including starbursts (Windhorst et al. 1985), Giga-Hertz Peaked Spectrum (GPS) sources and Compact Steep Spectrum (CSS) sources [see e.g. O'Dea (1998) for a review]. However one population
is thought to dominate, the Doppler boosted sources (e.g. Scheuer 1987). These are
preferentially selected in high-frequency samples because the
superposition of many synchrotron self-absorbed spectra along our
line-of-sight results in a flat-spectrum at high-radio frequencies.

As the radio emission is propagating along our line-of-sight in these
objects, the relativistic velocities associated with powerful radio
sources (e.g. Cohen et al. 1971) means that face-on radio
sources may undergo relativistic beaming which we see as a boost in
the flux.

From a statistical study of low-frequency and high-frequency selected
radio sources, Jackson \& Wall (1999) have shown that high-frequency
selected flat-spectrum sources have an opening angle within
$7^{\circ}$ of our line-of-sight. We note however that this value is
essentially a mean value and that both smaller or larger opening
angles are undoubtedly consistent with a Doppler boosting
paradigm. The opening angles may also depend on the intrinsic radio
power of the source (e.g. Jackson \& Wall 1999), thus the level of
Doppler boosting in a sample of flat-spectrum sources may have a wide
distribution.

However, keeping these caveats in mind, we can estimate the amount
of Doppler boosting the average flat-spectrum sources will exhibit,
compared with the Doppler boosting of the average quasar, if the
maximum opening angle is known for each population. Following the
method of Jarvis \& Rawlings (2000), we take the maximum opening angle
for which we observe a radio source as a quasar to be $53^{\circ}$, as
derived from the quasar fraction in low-frequency selected samples
(Willott et al. 2000). Consequently, averaging over solid angle, the
mean opening angle of steep spectrum radio-loud quasars is $\sim
37^{\circ}$ .

The boosting of the radio flux increases as
$\Gamma^{2}$:

\begin{equation}
\Gamma = \gamma^{-1} (1-\beta \cos \theta)^{-1} ,
\end{equation}

\noindent
where $\gamma$ is the Lorentz factor, $\beta=v/c$ and $\theta$ is the
angle between the radio jet and the line of sight. 
Therefore, adopting the conservative approach of 
substituting $\theta_{\rm flat} = 7^{\circ}$ for the flat-spectrum 
sources (many of the sources will have $\theta < 7^{\circ}$) and 
$\theta_{\rm steep} = 37^{\circ}$ for the steep spectrum quasar
population, we find that the Doppler
boosting factor is $\Gamma_{\rm flat}^{2}/\Gamma_{\rm steep}^{2}
\gtsim 100$. Hence, the intrinsic radio luminosity of the flat-spectrum
population is of the order $\gtsim 100$-times fainter than the
intrinsic radio luminosity of the average steep-spectrum radio-loud quasar.

In Fig.~\ref{fig:radpowvbhm1} we plot radio luminosity versus
black-hole mass with both the original radio luminosities, without any
boosting correction, and the same objects with the radio luminosity
decreased by a factor of 100. It is worth noting that this
amount of Doppler boosting will inevitably mean that some of these
radio-loud quasars should in fact be classified as radio intermediate
(e.g. Lacy et al. 2001) and in some cases radio-quiet. If we adopt the
radio-loudness parameter ${\mathcal R} = f_{\nu}{\rm (5~GHz)}/f_{\nu}{\rm (5100~\AA)}$, then many sources drop below the value usually taken as
the divide between radio-loud and radio-quiet quasars of ${\mathcal R} = 10$.
As highlighted by OWW02, the radio loudness parameter 
${\mathcal R}$ also depends on the amount of synchrotron emission 
which extends into the optical waveband. However, even taking this
into consideration, it is still likely that some of the sources will
now lie in the radio-quiet regime.  It is clear from 
Fig.~\ref{fig:radpowvbhm1} that following the correction for Doppler 
boosting the vast majority of the flat-spectrum
sources now lie within the $L_{\rm rad} - M_{bh}$ envelope of the quasar population
suggested by Dunlop et al. (2002).

This evidence is in itself enough to account for the major discrepancy
between the results of OWW02 and previous work. However, in this Section have
have only applied an average Doppler boosting correction factor to the
flat-spectrum sample as a whole. Obviously this average correction
factor will constitute an overestimate, or underestimate, depending
on the orientation of each individual object. This issue is considered
further in Section \ref{sec:reanalysis}. In the next Section
we proceed to consider the likely effect upon the estimated black-hole
masses of the flat-spectrum quasars due to their inclination close to
the line of sight.

\section{The geometry of the broad-line region}\label{sec:reanalysis2}

An indication of when a quasar is `misaligned' may also come
from the FWHM of the Balmer broad lines. The naive assumption is that
narrow ($\leq 4000$~\kmps) broad lines imply black holes of lower
mass. However, there is a wealth of evidence in the literature which
supports the view that the BLR has a disk-like geometry,
at least for radio-loud sources (e.g. Wills \& Browne 1986; 
Brotherton 1996; Vestergaard, Wilkes \& Barthel 2000). Using a sample
of 60 radio-loud quasars, Brotherton (1996) showed that the
radio core-to-lobe ratio, an indicator of radio source orientation, is
strongly correlated with the width of the broad H$\beta$ emission
line. The main implication being that pole-on radio sources 
have narrower broad lines than their misaligned
counterparts. Indeed, McLure \& Dunlop (2002) have shown that the 
H$\beta$ FWHM distribution of a sample of 72 AGN, both radio-loud and
radio-quiet, in the redshift interval $0.1<z<0.5$ can be naturally 
reproduced by considering the inclination effects expected if the BLR
has a flattened disk-like geometry.

If we now apply these results to the flat-spectrum sample considered
in this paper, the consequence is that the majority of the
measurements of $M_{bh}$ are essentially lower limits. Indeed, if
the BLR in these flat-spectrum sources are orientated within $\sim
7^{\circ}$ of the line of sight, then the black hole masses may be
underestimated by a factor $\gtsim 20$ (see e.g. McLure \& Dunlop
2002). 

However, here we adopt a conservative approach and use a low-frequency
radio selected quasar survey to predict what the mean FWHM of the 
broad Balmer lines should be, given no spectral-index 
selection criteria. We use the quasars from the Molonglo Quasar 
sample (MQS; Kapahi et al. 1998) for which line-width measurements 
are available in the literature (Baker et al. 1999). The mean FWHM of 
the H$\beta$ line in the MQS is $\approx 7000$~km~s$^{-1}$. In
contrast, the mean FWHM of the Balmer lines in the OWW02 flat-spectrum sample 
is $\sim 3500$~km~s$^{-1}$. We therefore choose to adopt 
a correction factor of two for the flat-spectrum FWHMs to compensate 
for orientation effects. Given that $M_{bh} \propto {\rm FWHM^{2}}$, this increases the black-hole mass estimates for 
the flat-spectrum sample by a factor of four.  The predicted position
of the flat-spectrum quasars on the $M_{bh}-L_{\rm rad}$ plane 
after application of the inclination correction is shown 
in Fig.~\ref{fig:radpowvbhm}, from which it can be seen that the
flat-spectrum quasars are now even more consistent with the upper and
lower radio power envelopes suggested by Dunlop et al. (2002).

\section{Radio spectral index and orientation}\label{sec:reanalysis}

Radio spectral index may be used to gain information on the
orientation of a radio source. As stated earlier, flat-spectrum
sources ($\alpha \sim 0$) are generally assumed to arise because of
the superposition of many optically thick regions along our
line-of-sight, whereas the emission from steeper-spectrum sources is
usually associated with optically thin radio lobes.

Therefore, although not explicitly applicable on a source-by-source
basis, there should be a general trend for steeper spectrum sources to
be orientated with the jets pointing away from our line-of-sight and
for the flatter-spectrum sources to be orientated with their jets
beamed along our line-of-sight.

Therefore, it interesting to note that the three sources which lie
closest to the lower envelope in Fig.~\ref{fig:radpowvbhm1} are sources 
with spectral indices steeper than $\alpha = 0.5$, and 
therefore should not strictly be in the flat-spectrum sample. 
Furthermore, in agreement with the inclination arguments outlined
above, these three sources also have the largest H$\beta$ FWHMs in the
OWW02 sample. If these three sources are in reality not aligned with 
their jets pointing within $7^{\circ}$ of our line of sight, then 
the correction for Doppler boosting could be significantly less than
the 
value of $\sim 100$ assumed throughout. Whereas, the sources with 
the narrowest broad lines in the OWW02 sample may be those 
in which the jet is pointing very close to our line-of-sight, and will
therefore have had their Doppler boosting correction factor 
underestimated. Together, these two factors work to
push the pole-on sources (i.e. those with $\theta < 7^{\circ}$) toward the
bottom right of Fig.~\ref{fig:radpowvbhm}, and the misaligned sources
(with $\theta > 7^{\circ}$) toward the top left of
Fig.~\ref{fig:radpowvbhm}. It is possible therefore that if 
each of the flat-spectrum quasars
could be corrected for Doppler boosting and BLR inclination on an
object-by-object basis, then they may actually be consistent with the
steep dependence of radio power on black-hole mass suggested by
previous studies (e.g. Franceschini et al. 1999; Laor 2000; Lacy et
al. 2001; Dunlop et al. 2002). 

\section{Conclusions}\label{sec:implications} 
We have re-analysed the data of Oshlack et al. (2002) on a sample of
flat-spectrum radio-loud quasars. Contrary to their conclusions we
find that, by correcting for the effects of inclination upon both the
radio luminosity and estimated black-hole mass, the black holes
harboured by intrinsically powerful flat-spectrum quasars are of
comparable mass to those found in other quasars of similar {\it
intrinsic} radio luminosity, i.e. $M_{bh} > 10^{8}$~M$_{\circ}$.

We also find that although many of the flat-spectrum quasars
occupy the region of intrinsic radio luminosity comparable to the FRII
radio sources (Fanaroff \& Riley 1974) found in low-frequency selected
radio surveys, some of the sources may occupy the
lower-luminosity regime of radio-intermediate and radio-quiet quasars.
Therefore, we conclude that by consideration of source inclination and
intrinsic radio power, flat-spectrum quasars may well be consistent with the $L_{\rm rad} \propto
M_{bh}^{2.5}$ relation found in previous studies.

Further
work is obviously essential to make firm statements regarding the
black-hole masses in flat-spectrum radio-loud quasars. 
This may be achieved by utilizing the bulge luminosity versus black-hole
mass correlation to determine the black-hole mass independent of any
orientation biases, and this is investigated in a 
subsequent paper (Jarvis, McLure \& Rawlings in prep.).

\section*{ACKNOWLEDGMENTS} 
MJJ acknowledges the support of the European Community Research and
Training Network "The Physics of the Intergalactic Medium". RJM
acknowledges PPARC funding. This
research has made use of the NASA/IPAC Extragalactic Database (NED)
which is operated by the Jet Propulsion Laboratory, California
Institute of Technology, under contract with the National Aeronautics
and Space Administration.

{} 

\end{document}